\documentclass[english,letterpaper,aps,prl,letterpaper,twocolumn,showpacs,floatfix,groupedaddress,superscriptaddress]{revtex4}
\usepackage{mathptmx}
\usepackage[T1]{fontenc}
\usepackage[latin9]{inputenc}
\usepackage{subfigure}
\usepackage{amsmath}
\usepackage{graphicx}
\usepackage{graphics}
\usepackage{setspace}
\usepackage{amssymb}

\makeatletter

\usepackage{mathptmx}
\makeatletter

\usepackage{mathptmx}

\usepackage{citesort}



\makeatother

\makeatother

\usepackage{babel}

\begin{document}

\title{Time-Dependent Density Functional Theory for Open Quantum Systems
with Unitary Propagation}

\author{Joel Yuen-Zhou}

\affiliation{Department of Chemistry and Chemical Biology, Harvard University,
12 Oxford Street, 02138, Cambridge, MA}

\author{David G. Tempel}

\affiliation{Department of Physics, Harvard University, 17 Oxford Street, 02138,
Cambridge, MA}

\author{César A. Rodríguez-Rosario}

\affiliation{Department of Chemistry and Chemical Biology, Harvard University,
12 Oxford Street, 02138, Cambridge, MA}

\author{Alán Aspuru-Guzik}

\affiliation{Department of Chemistry and Chemical Biology, Harvard University,
12 Oxford Street, 02138, Cambridge, MA}

\email{aspuru@chemistry.harvard.edu}

\begin{abstract}
We extend the Runge-Gross theorem for a very general class of Markovian
and non-Markovian open quantum systems under weak assumptions about
the nature of the bath and its coupling to the system. We show that
for Kohn-Sham (KS) Time-Dependent Density Functional Theory, it is
possible to rigorously include the effects of the environment within
a \emph{bath functional }in the KS potential, thus placing the interactions
between the particles of the system and the coupling to the environment
on the same footing. A Markovian bath functional inspired by the theory
of nonlinear Schrödinger equations is suggested, which can be readily
implemented in currently existing real-time codes. Finally, calculations
on a helium model system are presented. 
\end{abstract}


\maketitle
Current advances in the manipulation and control of nanoscale systems
allow for an unprecedented opportunity to probe the non-equilibrium
dynamics of a wide variety of condensed matter systems on a broad
range of timescales \citep{neuhauserscience,AlessandroTroisi09042007,E.Goulielmakis06202008}.
Serious effort is therefore required for the development of tractable
theoretical methods that can shed some light on many-body dynamics
without directly solving the time-dependent Schrödinger equation (TDSE)
for an object composed of many particles. One of the most promising
methods in this regard is Time-Dependent Density Functional Theory 
(TD-DFT) \citep{rungegross,tddft-review-burke,tddft-review-marques},
which is formally equivalent to the TDSE, but is based on the particle
density rather than the wavefunction.

Recently, there has been a considerable interest in developing an
Open Quantum Systems (OQS) formalism for TD-DFT, where the number
of particles in the system remains fixed, but there is energy exchange
with an environment \citep{Burke05a,DiVentra07a,burke-calculations,ullrich-vignale,r_gebauer_quantum_2005,dagosta-vignale,zheng:195127}.
This effort allows for the description of particle transfer within
the system, spontaneous decay, inelastic scattering, and many other
ubiquitous relaxation and dephasing phenomena. For a Markovian equation
of the Lindblad form, Burke, Car, and Gebauer (BCG) proved that a
statement analogous to the Runge-Gross (RG) theorem holds, namely,
that there is a one-to-one correspondence between the time-dependent
particle density and the external scalar potential provided that the
particle-particle interaction, initial quantum state, and the bath
jump-operators remain fixed \citep{Burke05a}. To place their result
in a practical context, BCG assumed the existence of a Kohn-Sham (KS)
scheme in order to carry out their calculations. In the mentioned
procedure, an artificial non-interacting open system, the so-called
KS system, evolves under an effective KS potential and is expected
to reproduce the particle density of the original system \citep{kohnsham}.
By virtue of their theorem, any observable is a functional of the
particle density, so in principle, the KS system contains all the
information about the observables of the original system. The question
of whether or not such a non-interacting KS system exists is not obvious,
but clearly crucial for KS theory, and it is known as the non-interacting
\emph{V-representability} problem \citep{vanLeeuwen1999,prbvignale}.
We note that non-interacting \emph{V}-representability in the context
of BCG's formalism has been assumed, but not formally proven.

In the case of closed systems, Van Leeuwen has proved that it is in
fact possible to reproduce the particle density of a many-body interacting
system with an effective KS potential acting on an auxiliary system
with no particle-particle interactions \citep{vanLeeuwen1999}. This
KS potential is unique, and in general, expected to show a nonlinear
and nonlocal funcional dependence on the history of the particle density
\citep{maitramemory}. Intuitively, we can argue that in the KS system 
we formally give up the linearity of the many-body equation of motion
for a nonlinear surrogate which, nevertheless, is an effective single-particle
equation. With this in mind, a natural question to ask is: Just as
with the particle-particle interactions, can we subsume the coupling
between the system and the bath into an additional nonlinearity of
the density in the effective KS potential? In the next paragraphs,
we report that this is indeed the case.

Consider an $N$-particle open quantum system described by a time-dependent
density matrix $\rho(t)$ which, in the position representation, is
a function of $6N$ coordinates and time $t$. The most general equation
of motion for an open quantum system is a master equation of the form 
(atomic units used throughout) \citep{breuer02a},

\begin{equation}
\dot{\rho}(t)=-i[\hat{H}(t),\rho(t)]+\int_{0}^{t}\mathcal{K}(t,t')\rho(t')dt'+\mathcal{T}(t).\label{eq:eqmotion}\end{equation}
Here, $\hat{H}(t)=\sum_{i}\left[\frac{|\hat{\vec{p}}_{i}|^{2}}{2m}+V(\hat{\vec{r}}_{i},t)\right]+\sum_{i<j}U(\hat{\vec{r}}_{i},\hat{\vec{r}}_{j})$
is the generator of the unitary piece of the evolution. In general,
$\hat{H}(t)$ is an effective renormalized Hamiltonian of the system
due to its interaction with the bath, where $U(\vec{r}_{i},\vec{r}_{j})$
is a symmetric pairwise interaction potential and $V(\vec{r},t)$
an external scalar potential. Finally, $\mathcal{K}(t,t')$ is a memory
kernel which describes the non-unitary effects of the bath on the
evolution of the system, and $\mathcal{T}(t)$ is an inhomogeneous
term which is present only if there are initial correlations between
the system and the bath. Quite generally, $\mathcal{K}(t,t')$ and
$\mathcal{T}(t)$ may be functions of $V(\vec{r},t)$, such as in
the case of a strong laser field interacting with a molecule in condensed
phase \citep{meier:3365}, whereas $\mathcal{T}(t)$ may also depend
on the initial state $\rho(0)$. 

Furthermore, for notation, we define the operators that measure the
particle density as $\hat{n}(\vec{r})=\sum_{i}\delta(\vec{r}-\hat{\vec{r}}_{i})$
and the current density as $\hat{\vec{j}}(\vec{r})=\frac{1}{2}\sum_{i}\{\delta(\vec{r}-\hat{r}_{i}),\hat{v}_{i}\}$.
We are now ready to state a theorem.

\emph{Theorem.-} Let the \emph{original} system be described by the
density matrix $\rho(t)$ which, starting as $\rho(0)$, evolves according
to Eq. (\ref{eq:eqmotion}). Consider an \emph{auxiliary} system associated
with the density matrix $\rho'(t)$ and initial state $\rho'(0)$,
which is governed by the equation:
\begin{equation}
\dot{\rho}'(t)=-i[\hat{H}'(t),\rho'(t)]+\int_{0}^{t}dt'\mathcal{K}'(t,t')\rho'(t')+\mathcal{T}'(t),\label{eq:lindblad prime}\end{equation}
where the functional forms of $\mathcal{K}'(t)$ and $\mathcal{T}'(t)$
are given, and its Hamiltonian reads as $\hat{H}'(t)=\sum_{i}\left[\frac{|\hat{\vec{p}}_{i}|^{2}}{2m}+V'(\hat{\vec{r}}_{i},t)\right]+\sum_{i<j}U'(\hat{\vec{r}}_{i},\hat{\vec{r}}_{j})$,
where $U'(\vec{r}_{i},\vec{r}_{j})$ is also given. Under mild conditions,
there exists an external potential $V'(\vec{r},t)$ that drives the
auxiliary system in such a way that the particle densities in the
original and the auxiliary systems are the same at every point in
time and space, i.e., $\langle\hat{n}(\vec{r})\rangle_{t}'=\langle\hat{n}(\vec{r})\rangle_{t}$.
This statement is true provided that $\rho'(0)$ guarantees that $\langle\hat{n}(\vec{r})\rangle_{t=0}'=\langle\hat{n}(\vec{r})\rangle_{t=0}$
and $\langle\dot{\hat{n}}(\vec{r})\rangle_{t=0}'=\langle\dot{\hat{n}}(\vec{r})\rangle_{t=0}$. 

\emph{Proof.-} We use similar techniques to the ones employed by van
Leeuwen \citep{vanLeeuwen1999} and Vignale \citep{prbvignale}. The
detailed steps of a related derivation may be found in \citep{YuenZhou2009PCCP}.
First, by using Eq. (\ref{eq:eqmotion}), we can find the equation
of motion for the second derivative of the particle density of the
original system with respect to time:
\begin{eqnarray}\label{eq:evolution density}
\langle\ddot{\hat{n}}(\vec{r})\rangle_{t} & = & \vec{\nabla}\cdot[\langle\hat{n}(\vec{r})\rangle_{t}\vec{\nabla}V(\vec{r},t)/m+\vec{\mathcal{D}}(\vec{r},t)+\nonumber \\
 &  & \vec{\mathcal{F}}(\vec{r},t)/m+\vec{\mathcal{G}}(\vec{r},t)]+\mathcal{J}(\vec{r},t).\end{eqnarray}
Here, $\vec{\nabla}V(\vec{r},t)$ is proportional to the external
electric field, $\vec{\mathcal{D}}(\vec{r},t)=-\frac{1}{4}\sum_{\alpha,\beta}\hat{\beta}\frac{\partial}{\partial\alpha}\left\langle \sum_{i}\{\hat{v}_{i\alpha},\{\hat{v}_{i\beta},\delta(\vec{r}-\hat{\vec{r}}_{i})\}\}\right\rangle $
is the divergence of the stress tensor, where $\alpha,\beta=x,y,z$,
$\vec{\mathcal{F}}(\vec{r},t)$ is the internal force density caused
by the pairwise potential $\vec{\mathcal{F}}(\vec{r},t)=-\langle\sum_{i}\delta(\vec{r}-\hat{\vec{r}}_{i})\sum_{j\neq i}\vec{\nabla}_{\vec{r}_{i}}U(\vec{r}_{i}-\vec{r}_{j})\rangle$,
and $\vec{\mathcal{G}}(\vec{r},t)=Tr\{\hat{\vec{j}}(\vec{r})(\int_{0}^{t}dt'\mathcal{K}(t,t')\rho(t')+\mathcal{T}(t))\}$
and $\mathcal{J}(\vec{r},t)=\frac{\partial}{\partial t}Tr\{\hat{n}(\vec{r})\left(\int_{0}^{t}dt'\mathcal{K}(t,t')\rho(t')+\mathcal{T}(t)\right)\}$
are terms which arise due to the coupling to the bath. Similarly,
by employing Eq. (\ref{eq:lindblad prime}), it is possible to derive
an equivalent equation for $\langle\ddot{\hat{n}}(\vec{r})\rangle_{t}'$,
where the variables in Eq. (\ref{eq:evolution density}) are substituted
by their primed analogues. If we subtract these two equations and
eliminate the variable $\langle\hat{n}(\vec{r},t)\rangle'$ with the
restriction $\langle\hat{n}(\vec{r})\rangle_{t}'=\langle\hat{n}(\vec{r})\rangle_{t}$,
we obtain an identity with time-dependent parameters that can be Taylor
expanded about $t=0$. Denoting the Taylor expansion coefficients
by $O_{k}\equiv\frac{1}{k!}\frac{\partial^{k}O(\vec{r},t)}{\partial t^{k}}|_{t=0},$ we
collect the terms of order $t^{l}$, and arrive at the expression:

\begin{align}
-\vec{\nabla}\cdot(n_{0}(\vec{r})\vec{\nabla}(V'_{l}(\vec{r}))) & =\nonumber \\
-\vec{\nabla}\cdot(m\vec{\mathcal{D}}'_{l}(\vec{r})+\vec{\mathcal{F}}'_{l}(\vec{r})+m\vec{\mathcal{G}}'_{l}(\vec{r}))+m\mathcal{J}'_{l}(\vec{r})\nonumber \\
+\vec{\nabla}\cdot(m\vec{\mathcal{D}}{}_{l}(\vec{r})+\vec{\mathcal{F}}{}_{l}(\vec{r})+m\vec{\mathcal{G}}{}_{l}(\vec{r}))-m\mathcal{J}{}_{l}(\vec{r})\nonumber \\
-\vec{\nabla}\cdot(n_{0}(\vec{r})\vec{\nabla}(V{}_{l}(\vec{r})))+\vec{\nabla}\cdot\left(\sum_{k=1}^{l}n_{k}(\vec{r})\vec{\nabla}(V'_{l-k}-V{}_{l-k}(\vec{r}))\right) & .\label{eq:recursion3}\end{align}
We now make a claim: If the right hand side of Eq. (\ref{eq:recursion3})
contains no coefficients $V'_{k}(\vec{r})$ for $k\geq l$, it can
be regarded as a recursion relation to construct $V'_{l}$ from the
lower order coefficients $V'_{k}(\vec{r})$ for $0\leq k<l$. This
would imply that each coefficient can be \emph{uniquely} solved recursively
upon the specification of a boundary condition, which we can conveniently
set to $V'_{l}(\vec{r})\to0$ as $|\vec{r}|\to\infty$, for all $l$.
Finally, the explicit construction of $V'(\vec{r},t)$ through its
Taylor coefficients, $V'(\vec{r},t)=\sum_{k}V'_{k}(\vec{r})t^{k}$,
proves the theorem. 

If $\mathcal{K}'(t,t')$ and $\mathcal{T}'(t)$ do not depend explicitly
on $V'(\vec{r},t)$, the claim can be systematically shown \citep{YuenZhou2009PCCP}.
Otherwise, $\mathcal{J}'_{l}(\vec{r})$ can depend at most on 
${\mathcal{K}'}_{l}^{t=t'}\equiv\frac{1}{l!}\frac{\partial^{l}{\mathcal{K}'}(t,t)}{\partial t^{l}}|_{t=0}$
(the integral terms $\int_{0}^{t}dt(\cdot)$ naturally vanish at $t=0$)
and $\mathcal{T}'_{l}=\frac{1}{l!}\frac{\partial^{l}\mathcal{T}'(t)}{\partial t^{l}}|_{t=0}$.
General expressions derived with projection-operator methods \citep{meier:3365}
can be used to formally show that ${\mathcal{K}'}_{l}^{t=t'}$ should
depend at most on $V'_{l-1}(\vec{r})$, which supports our claim.
This fact can be interpreted in very physical terms: the action of
the external field $V'(\vec{r},t)$ on the system is local in time
through the unitary piece of the master equation. The effects of $V'(\vec{r},t)$
on the system leak out to the bath and return as memory effects through
the memory kernel only at times $t''$ \emph{strictly} later than
$t$. In other words, $\mathcal{K}'(t,t)$ can depend on $V'(\vec{r},t')$
for $t'<t$, but should not depend on the instantaneous $V'(\vec{r},t)$. 

A similar conclusion may not be made for arbitrary $\mathcal{T}_{l}^\prime$
terms, since at $t=0$, the initial correlations between the system
and the bath may depend on $V(\vec{r},0)$, and $\mathcal{T}_{l}^\prime$
could depend on $V'_{l}(\vec{r})$. However, as long as $\mathcal{T}_{l}^\prime$
depends at most on $V'_{l-1}(\vec{r})$, the claim and the theorem
will necessarily hold. This is the only warning of the proof, and
this requirement can be checked on a case by case basis, but it is
easily guaranteed in the case of initial factorizable conditions between
the system and the bath, or if the inhomogeneity is $V$-independent,
which occurs if the external field is weak or if the bath is Markovian.
$\square$

Several important corollaries hold from the theorem. If $\rho'(0)=\rho(0)$,
$U'(\vec{r}_{i},\vec{r}_{j})=U(\vec{r}_{i},\vec{r}_{j})$, $\mathcal{K}'(t,t')=\mathcal{K}(t,t')$,
and $\mathcal{T}'(t)=\mathcal{T}(t)$, then Eq. (\ref{eq:recursion3})
reads: $-\vec{\nabla}\cdot(n_{0}\vec{\nabla}(V'_{l-k}-V{}_{l-k}))=\vec{\nabla}\cdot\left(\sum_{k=1}^{l}n_{k}\vec{\nabla}(V'_{l-k}-V{}_{l-k})\right)$,
which means that $V_{l}'=V_{l}$ for all $l$. This allows for an
extension of the RG theorem to a large class of OQS: For fixed initial
state, interparticle potential, memory kernel and inhomogeneity, there
is a one to one map between particle densities and scalar potentials.
This statement allows us to regard the time-dependent particle density
as a fundamental variable just as the time-dependent density matrix.
For Markovian equations of the Lindblad form, this reduces to the
result proven by GCB.

The theorem also justifies the KS scheme of BCG and its generalization
to a wide range of OQS, namely, that it is possible to choose an auxiliary
open system with no particle-particle interactions, $U'(\vec{r}_{i},\vec{r}_{j})=0$,
to reproduce the same particle density as the original system. However,
we want to take a different approach on the subject and make the observation
that the proof also allows us to consider the case where $U'(\vec{r}_{i},\vec{r}_{j})=\mathcal{K}'(t,t')=\mathcal{T}'(t)=0$,
that is, a KS system that evolves unitarily as if it were a driven
closed system, but still reproduces the particle density of the original
open system that interacts with the bath and evolves through a non-unitary
equation of motion. Therefore, we have rigorously justified the intuition
hinted at the beginning of the letter, that is, the possibility to
conceive of a KS system where we subsume the effects of the bath in
an additional term in the KS potential 
\footnote{We clarify that other observables besides the particle density many
not be the same in the driven closed KS system when compared to the
ones corresponding to the open original system. On the other hand,
due to the RG-like statement we have derived, any observable is a
\emph{functional} of the particle density, which is in principle the
same in both original and KS systems.}.
In this new KS theory, we shall rewrite the KS potential as $V'=V+V_{H}+V_{xc}+V_{bath}$,
where $V$ is the original external potential, $V_{H}(\vec{r},t)=\int d^{3}r'\frac{\langle\hat{n}(\vec{r}')\rangle_{t}}{|\vec{r}-\vec{r}'|}$
is the Hartree term, $V_{xc}$ is a standard approximation to the
exchange-correlation (xc) term due to the many-body effects within
the system, such as an adiabatic functional \citep{baerprevalence},
and finally, $V_{bath}$ is the new term due to the bath, which includes
additional correlations on the particles of the system, and which
we expect to be non-adiabatic. Finally, we must discuss the feasability
of the initial conditions for our KS scheme. It is always possible
to propose a pure state single Slater determinant $\tilde{\psi}'(0)=\frac{1}{\sqrt{N!}}\det[\phi_{i}(\vec{r}_{j})]$
which satisfies the restriction $\langle\tilde{\psi}'(0)|\hat{n}(\vec{r})|\tilde{\psi}'(0)\rangle=\langle\hat{n}(\vec{r})\rangle_{t=0}$
by employing the Harriman construction \citep{Harriman}. By defining
a new state $\psi'(0)=\frac{1}{\sqrt{N!}}\det[\phi_{i}(\vec{r}_{j})e^{i\alpha_{i}(\vec{r}_{j})}]$,
the set of phases $\{\alpha_{i}\}$ can be chosen with considerable
freedom in order to satisfy $\frac{\partial}{\partial t}\langle\psi'(t)|\hat{n}(\vec{r})|\psi'(t)\rangle|_{t=0}=-\vec{\nabla}\cdot\left(\sum_{i}|\phi_{i}(\vec{r})|^{2}\vec{\nabla}(-i\arg(\phi_{i}(\vec{r})+\alpha_{i}(\vec{r}))\right)=\langle\dot{\hat{n}}(\vec{r})\rangle_{t=0}$,
in which case, we can choose $\psi'(0)$ as the initial KS wavefunction,
or equivalently, $\rho'(0)=|\psi'(0)\rangle\langle\psi'(0)|$ as the
initial KS density matrix. Note that this argument is irrespective
of the purity of the initial state of the original system. 

\emph{Model system and suggestion of {}``bath'' functional.- }We
refer the reader to Ref. \citep{YuenZhou2009PCCP}, which reports
a numerical study that constructs the KS potential $V'$ for a harmonic
oscillator model coupled to a heat bath. In this letter, we will be
concerned with the study of a model system, namely, a 1-d helium atom
\citep{thiele:153004,harabati:084104} coupled to a heat bath. We
write the total system-bath Hamiltonian as $H_{T}=H_{S}+H_{SB}+H_{B}$.
$H_{S}=\sum_{i=1}^{2}\left(P_{i}^{2}/2+V(X_{i},t)\right)+W(X_{1}-X_{2})$
describes the helium atom, with $X_{i}$ and $P_{i}$ denoting the
positions and momenta of the electrons, $W(X)=e^{2}/\sqrt{X^{2}+1}$
being a soft-Coulomb potential, and $V(X)=-2W(X)$ the external potential,
which in this case is only due to the nucleus. $H_{B}+H_{SB}=\frac{1}{2}\sum_{j}m_{j}\left[\dot{x}_{j}^{2}+\sum_{i}\omega_{j}^{2}\left(x_{j}-\frac{c_{j}}{m_{j}\omega_{j}^{2}}X_{i}\right)^{2}\right]$
corresponds to a harmonic bath with bilinear coupling to the positions
of the electrons. We assume that the bath is an infinite set and its
distribution of couplings can be approximated by a continuous Ohmic
spectral density, $J(\omega)=\sum_{j}\frac{c_{j}^{2}}{2m_{j}\omega_{j}}\delta(\omega_{j}-\omega)=\theta(\omega)\frac{\xi_{0}}{2}\omega e^{-\omega/\omega_{c}}$,
 where $\theta(\omega)$ is the step function, $\xi_{0}$ is the intensity
 of the coupling, and $\omega_{c}$ is a cutoff frequency for the bath
 modes. From a computational point of view, the dynamics of the composite
 system-bath object is intractable. Since the emphasis is on the system,
 and not on the bath, we take an OQS approach: For weak coupling $\xi_{0}$
 and large $\omega_{c}$, the Born-Markov approximation is justified,
 and it is straightforward to obtain a memoryless master equation of
 the Lindblad form for the \emph{system}. At zero temperature ($T=0$),
 it reads,

 \begin{equation}
 \dot{\rho}(t)=-i[\tilde{H}_{S},\rho]-\frac{\gamma}{2}(L^{\dagger}L\rho+\rho L^{\dagger}L-2L\rho L^{\dagger}),\label{eq:lindblad helium}\end{equation}
 where $\tilde{H}_{S}=H_{S}+\frac{\xi_{0}\omega_{c}}{2}(x^{2}+y^{2})$
 is a renormalized Hamiltonian due to coupling to the bath. We denote
 $|g\rangle$ and $|e\rangle$ to be the ground and first singlet excited
 states of $\tilde{H}_{S}$ respectively, so that the jump operators
 $L$ can be expressed in the form $L=|g\rangle\langle e|$. $L$ promotes
 quantum jumps from $|e\rangle$ to $|g\rangle$. The rate of these
 transitions is captured by $\gamma=2\pi|\langle e|\mu|g\rangle|^{2}J(\omega_{eg})$,
 where $\mu=\sum_{i=1}^{2}X_{i}$ is the dipole operator.

 We proceed to derive a bath functional which could be used in the
 KS theory for TD-DFT applied to systems interacting with a Markovian
 bath, just like our model system. For a single particle, Kostin \citep{kostin:3589}
 has previously constructed a dissipative nonlinear Schrödinger equation,
 where $i\frac{\partial\psi}{\partial t}=H\psi$, for which the Hamiltonian
 in 1-D reads $H=\frac{p^{2}}{2M}+V+V_{bath}$, with the bath potential
 being given by $V_{bath}(X,t)=\frac{\lambda}{2i}\ln\left(\frac{\psi(X,t)}{\psi^{*}(X,t)}\right)$.
 This equation of motion has the very interesting property that at
 the level of observables, it satisfies the Langevin equation at $T=0$,
 i.e., $\langle\dot{X}\rangle=\frac{\langle P\rangle}{M}$, $\langle\dot{P}\rangle=-\lambda\langle P\rangle-\langle\frac{\partial V(Z,t)}{\partial Z}\rangle$,
 as can easily be checked by direct substitution. The friction coefficient
 $\lambda$ may be obtained from a microscopic derivation of the Langevin
 equation, which in the case of a particle bilinearly coupled to an
 Ohmic bath of strength $\xi_{0}$, yields $\lambda=\pi\xi_{0}/2$.
 Furthermore, a quick inspection allows us to rewrite $V_{bath}$ as
 a functional of the particle density %
 \footnote{In 1-d, we can express the current as a functional of the particle
 density, $\langle\hat{j}(z)\rangle_{t}=-\int_{-\infty}^{z}\frac{\partial\langle\hat{n}(z')\rangle}{\partial t}dz'$,
 where we have assumed $\langle\hat{j}(\pm\infty)\rangle=0$. For more
 dimensions, this might not be possible, as we will explain, this is
 not a problem from a practical perspective in the KS propagation.%
 },\\
\begin{equation}
V_{bath}[\langle\hat{n}(X')\rangle_{t},\langle\hat{j}(X')\rangle_{t}](X,t)=\lambda\int_{-\infty}^{X}dX'\frac{\langle\hat{j}(X')\rangle_{t}}{\langle\hat{n}(X')\rangle_{t}}.\label{eq:functional}\end{equation}
 For more than one particle, this identification is not formally possible,
but regardless, we shall heuristically assume it as our \emph{Markovian}
\emph{bath functional} (MBF) 
\footnote{Eq. (\ref{eq:functional}) will not be able to cause transitions from
eigenstates, which have the property $\langle\hat{j}(X')\rangle_{t}=0$.
For these cases, the inclusion of a small penalty functional $\kappa(\langle\hat{n}(X)\rangle_{t}-\langle\hat{n}(X)\rangle_{target})$
to $V_{bath}$ guarantees the proper evolution of the KS system. Here,
$|\frac{\kappa}{\lambda}|\frac{1}{a_{0}}\ll1$, and $\langle\hat{n}(X)\rangle_{target}$
represents the final steady state particle density,%
}. Non-Markovian generalizations of Eq. (\ref{eq:functional}) may
be readily conceived starting from nonlinear Schrödinger equations
which reproduce the generalized Langevin equation for its observables.
Physically, this suggestion is very appealing: The dragging force
due to the MBF is proportional to $\frac{\langle\hat{j}(X)\rangle_{t}}{\langle\hat{n}(X)\rangle_{t}}$,
which is the velocity field. The coefficient $\lambda$ can be approximated
from the spectral density and conveniently scaled to reflect the many-body
coupling to the bath. From the single Slater determinant KS wavefunction,
$\psi_{KS}(t)=\frac{1}{\sqrt{N!}}\det[\phi_{i}(X_{j},t)]$, we can
express $V_{bath}(X,t)=\lambda\int_{c}^{z}dX'\frac{\sum_{i}|\phi_{i}(X',t)|^{2}\nabla\alpha_{i}(X',t)}{\sum_{i}|\phi_{i}(x,t)|^{2}}$,
where $\alpha_{i}=-i\arg(\phi_{i})$. 


In order to gain insight on the system of consideration, we performed
several calculations for which the results are summarized in Fig.
1. The initial state of the helium atom was taken to be the pure state
$\psi(0)=\frac{1}{\sqrt{2}}(|g\rangle+|e\rangle)$. We propagated
the system in real time with three different methods. For the first
method (black solid curve), we evolved the density matrix of helium
using the master Eq. (\ref{eq:lindblad helium}). We chose a spectral
density with values $\xi_{0}=0.01\, E_{h}$ and $\omega_{c}=10\, E_{h}$.
The real space eigenbasis of $\tilde{H}_{S}$ was obtained with the
OCTOPUS package \citep{octopus}, resulting on an energy gap $\Delta_{eg}=0.85\, E_{h}$
and a dipole moment $\langle e|\mu|g\rangle=1.1\, a.u$. The choice
of parameters justifies the Markovian conditions for the master equation.
The expected damped oscillations calculated with this method are shown
in the solid curve. The second method (red solid curve) was performed
to calibrate the parameter $\lambda$ in $V_{dis}$. We evolved the
time dependent Schrödinger equation with the effective Hamiltonian
$\tilde{H}_{S}+V_{dis}$ using the Suzuki-Trotter split operator method
\citep{kosloffpropagators}, where the many-body dynamics was computed
exactly via $\tilde{H}_{S}$, but the coupling to the bath entered
through the nonlinear dependence of $V_{dis}$ on $\langle\hat{n}(\vec{r})\rangle_{t}$.
We scanned several $\lambda$ parameters and found $\lambda=0.075\, E_{h}$
to reproduce the curve derived from (A) with high accuracy 
\footnote{From a microscopic derivation, it is possible to argue that an approximate
friction coefficient arising from the coupling of the bath to two
electronic coordinates could be $\lambda\approx\frac{\xi_{0}\pi}{2\sqrt{2}}=0.01\, E_{h}$, which differs from the optimized value. The difference between these
two values may be due to the lack of dependence of $\lambda$ on $\omega_{c}$.
A more systematic derivation of $\lambda$ and a detailed examination
of this problem will be addressed in future work.}. Finally, for the third method (black dotted curve) we carried out
a TD-DFT KS calculation with exact exchange and same dissipation rate
$\gamma$ as in B, that is, $V_{KS}=\frac{1}{2}V_{H}+V_{dis}$, with
$V_{H}(X,t)=\gamma\int dX'\frac{\langle\hat{n}(X')\rangle_{t}}{\sqrt{1+(X-X')^{2}}}$.
The result for this last method yields poor results with unphysical
Rabi-like oscillations. The latter are caused by the absence of correlations
caused by particle-particle interactions \citep{ruggenthaler:233001}.
Nevertheless, the oscillations decay on a similar timescale to the
other calculations, and reach a steady state due to the MBF. 


\begin{figure}
\begin{center}
\subfigure[]{\includegraphics[scale=0.6]{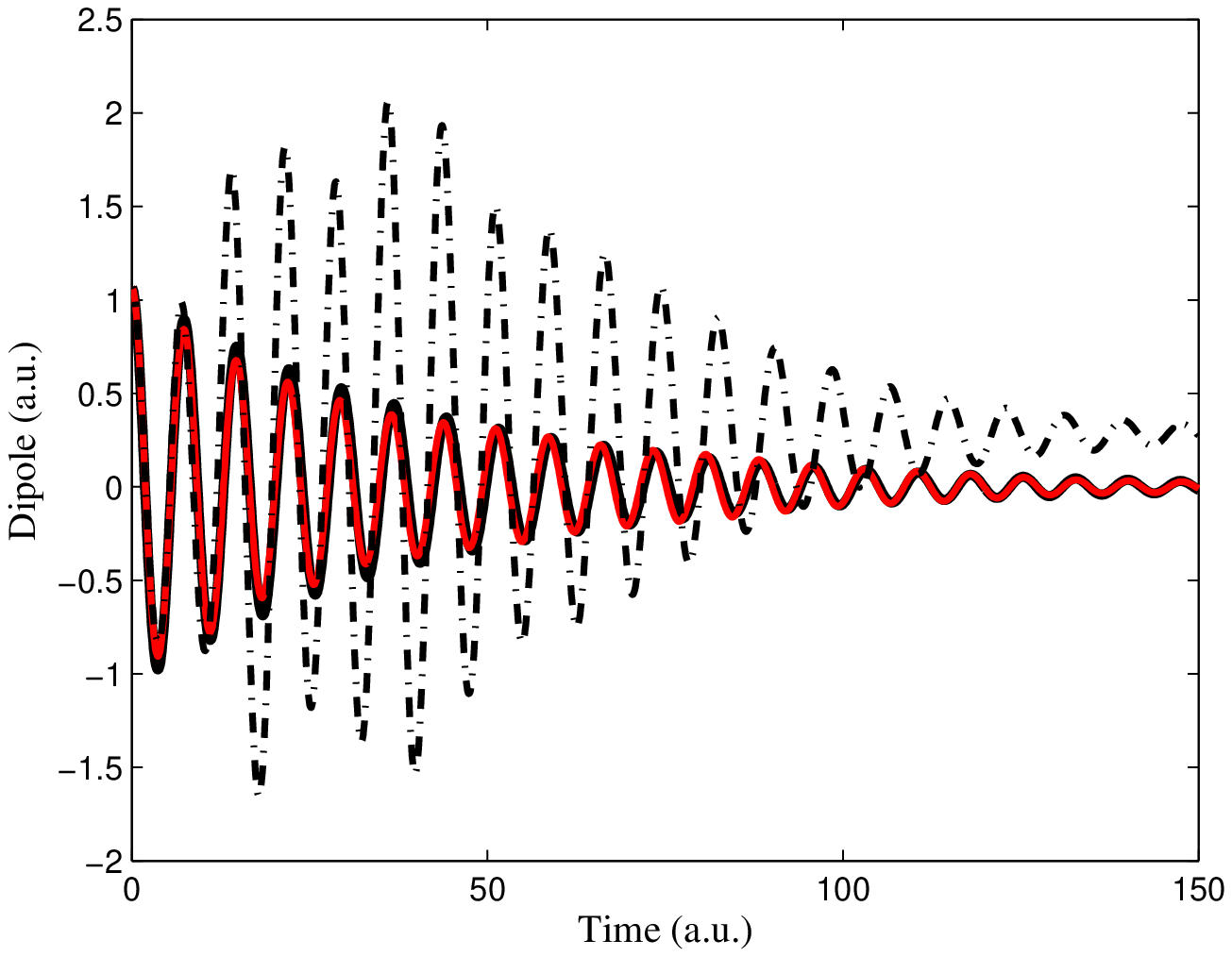}}
\par\end{center}

\begin{spacing}{0.7}
\begin{raggedright}
{\footnotesize FIG. 1. Evolution of the dipole moment of a helium
atom coupled to a heat bath. We present three different calculations:
The }\emph{\footnotesize black solid}{\footnotesize{} curve represents
the \char`\"{}exact\char`\"{} calculation using a master equation.
The }\emph{\footnotesize red solid}{\footnotesize{} curve is the propagation
of the exact many-body dynamics of helium plus the MBF. Finally, the
}\emph{\footnotesize dotted black}{\footnotesize{} is the TD-DFT calculation
with exact exchange and MBF. The last calculation yields poor results
due to the absence of correlations in the electron interactions. Details
of the calculations can be found in the text.}
\par\end{raggedright}\end{spacing}

\end{figure}


In summary, we have formally extended TD-DFT to a large class of OQS,
and rigurously showed the possibility of including the effects of
the bath on the dynamics of the system within a bath functional. The
latter enters into a TD-DFT calculation on the same footing as the
standard exchange-correlation functionals exclusively due to many-body
dynamics. We have suggested Eq. (\ref{eq:functional}) as the Markovian
bath functional which can be readily implemented in currently existing
TD-DFT codes. Future work must address the derivation of the friction
coefficient $\lambda$ from a more systematic procedure, the explicit
derivation of bath functionals for more complex memory kernels, and
finally, the role of fluctuations and finite temperatures in this
unitary propagation KS formalism. 

Stimulating discussions with many members of the Aspuru-Guzik group
are greatly acknowledged. J.Y.Z. also thanks the generous support
of Fundación M\'exico at Harvard and CONACYT. C.A.R. thanks the Mary-Fieser
Postdoctoral Fellowship program. This work was carried out under the
DARPA contract FA 9550-08-1-0285.


\end{document}